\documentstyle[aps,prb,epsf,epsfig,multicol]{revtex}
\begin{document}
\draft
\date{\today}
\title
{Macroscopic  anisotropy in superconductors with anisotropic gaps}
\author{V. G. Kogan }
\address{ Ames Laboratory DOE and Physics Department ISU, Ames, IA 
50011 }
\maketitle

\begin{abstract}
It is shown within the weak-coupling model that the macroscopic
superconducting anisotropy for materials with the gap varying on
the Fermi surface cannot be characterized by a single number, unlike the
case of clean materials with  isotropic gaps. For  clean uniaxial
materials, the anisotropy parameter $\gamma (T)$ defined as
 the ratio of London penetration depths, $\lambda_c/\lambda_{ab}$, is
evaluated for all $T$'s. Within the two-gap model of MgB$_2$,  $\gamma (T)$
is an increasing function of $T$.
\end{abstract}
\begin{multicols}{2}
 
  A remarkable confirmation for the observed two-gap
structure\cite{Bouquet,Szabo,Giubileo,Junod,Schmidt} of superconducting
MgB$_2$ came from solving the Eliashberg equations for the gap distribution
on the Fermi surface.\cite{Choi,Mazin} According to this, the gap on the
four Fermi surface sheets of this material has two sharp maxima:
$\Delta_1\approx 1.7\,$meV at the two   $\pi$-bands  and 
$\Delta_2\approx 7\,$meV at the two $\sigma$-bands. Within each of
these groups, the spread of the gap values is small, and the gaps can be
considered as constants, the ratio of which is nearly $T$ independent. In
this situation, a weak-coupling model with two gaps on two parts of the
Fermi surface may prove  useful in relating various macroscopic properties
of MgB$_2$. Starting with Ref. \onlinecite{SMW}, the two-band models were
studied by many, see, e.g., Ref. \onlinecite{mazin} and references therein.
The focus of this work is on the macroscopic superconducting anisotropy
$\gamma$. To a large extent, motivation for this work was to understand why
experiments on different samples of MgB$_2$ done with different techniques
yield widely varying values for  
$\gamma$.\cite{Lima,madison,Simon,BKC,Angst,Solog,Shinde} 

The anisotropic Ginzburg-Landau (GL) equations,  derived for clean
superconductors with an arbitrary gap anisotropy in the seminal work by
L.Gor'kov and  T.Melik-Barkhudarov,\cite{Gorkov} led to the commonly
used concept of a single parameter $\gamma$ defined   as  
$\xi_a/\xi_c\equiv \lambda_c/\lambda_a$ ($\xi$ is the coherence length,
$\lambda$ is the penetration depth, and $a,c$ are principal crystal
directions). Formally, this came out because the same mass 
tensor enters both the first GL equation which determines the anisotropy 
of $\xi$ (and of the upper critical fields $H_{c2}$) and the equation for
the current which defines the anisotropy of $\lambda$.  However,  it has
been  later shown by S.Pokrovsky and V.Pokrovsky in the work on the GL
equations for anisotropic gaps in the presence of impurities, that $\gamma$
in fact depends on the impurity scattering, i.e., it might be sample
dependent.\cite{PP} 
 
In the following   the near-$T_c$ result of Ref. \onlinecite{PP} is
reproduced using the Eilenberger formalism. Moreover, the ratio $
\lambda_c/\lambda_a$ for arbitrary temperatures $T$ is derived for the clean
case. It is shown that for MgB$_2$, $\lambda_c/\lambda_a$ should increase 
with increasing $T$, the result which calls for experimental
verification. \\ 


  We begin with the quasiclassical  version of the BCS theory  for
a general anisotropic Fermi surface: \cite{E} 
\begin{eqnarray}
{\bf v} {\bf \Pi}f&=&2\Delta g/\hbar  -2\omega f+(g\langle f\rangle
-f\langle g\rangle)/\tau\,,
\label{eil1}\\
-{\bf v} {\bf \Pi}^*f^+&=&2\Delta^* g/\hbar  -2\omega f^++(g\langle
f^+\rangle -f^+\langle g\rangle)/\tau,
\label{eil2}\\
g^2&=&1-ff^{+}\,, \label{eil3}\\
\Delta({\bf r},{\bf v})&=&2\pi TN(0) \sum_{\omega >0}^{\omega_D} \langle
V({\bf v},{\bf v}^{\prime\,}) f({\bf v}^{\prime\,},{\bf r},\omega)\rangle_{{\bf
v}^{\prime\,}}\,.   \label{eil4}\\
{\bf j}&=&-4\pi |e|N(0)T\,\, {\rm Im}\sum_{\omega >0}\langle {\bf
v}g\rangle\,.
\label{eil5}
\end{eqnarray}
Here ${\bf v}$ is the Fermi velocity, ${\bf \Pi} =\nabla +2\pi i{\bf 
A}/\phi_0$; $\Delta $ is the gap function,
$f({\bf r},{\bf v},\omega),\,\,  f^{+} $, and $g$ are Eilenberger Green's 
functions, $N(0)$ is the total density of states at the Fermi level per one
spin; $\hbar\omega=\pi T(2n+1)$ with an integer $n$. Further, $\tau$ is
the scattering  time on non-magnetic impurities and $\omega_D$ is the Debye 
frequency. The averages over the Fermi surface weighted with the local
density of states $
\propto 1/|{\bf v}|$ are defined as
\begin{equation}
\langle X \rangle = \int \frac{d^2{\bf k}_F}{(2\pi)^3\hbar N(0)|{\bf v}|}\,
\,X\,.
\label{<>}
\end{equation}
 
Commonly, the interaction $V$ is assumed factorizable,\cite{Kad}  
 $ V({\bf v},{\bf v}^{\prime\,})=V_0 \,\Omega({\bf v})\,\Omega({\bf
v}^{\prime\,})$, and one looks for   $\Delta (
{\bf r},T;{\bf v})=\Psi ({\bf r},T)\, \Omega({\bf v})$. Then, the
self-consistency Eq. (\ref{eil4}) takes the form:
 \begin{equation}
\Psi( {\bf r},T)=2\pi T N(0)V_0 \sum_{\omega >0}^{\omega_D} \langle
\Omega({\bf v} ) f({\bf v} ,{\bf r},\omega)\rangle \,.
\label{gap}
\end{equation}

The  function $\Omega({\bf v})$ can be normalized  by requiring that the
critical temperature $T_{c0}$ for the {\it clean} material ($\tau\to\infty$)
is given by the standard isotropic weak-coupling model with the effective
interaction $V_0$:  
\cite{Pokr}
\begin{equation}
 \langle \Omega^2 \rangle=1\,.  
\label{norm}
\end{equation}

 As usual, we incorporate  $T_{c0}$ in the Eilenberger
system using the identity 
\begin{equation}
\frac{ 1}{N(0) V_0}= \ln \frac{T}{T_{c0}}+2\pi T\sum_{\omega >0}^{\omega_D}
 \frac{ 1}{\hbar \omega} \,.
\label{1/NV}
\end{equation}
Substitute this in   Eq. (\ref{gap}) and replace $\omega_D$ with infinity
due to the fast convergence:
 \begin{equation}
\frac{\Psi }{2\pi T} \ln \frac{T_{c0}}{T}= \sum_{\omega
>0}^{\infty}\Big(\frac{\Psi}{\omega}-\langle \Omega \, f \rangle \Big)\,.
\label{gap1}
\end{equation}


{\it Effect of nonmagnetic impurities on $T_c$.} 
It is long known  that scattering by nonmagnetic impurities 
suppress $T_c$ provided the gap is weakly anisotropic.\cite{Kad,Hoh}  The 
suppression is readily obtained from Eilenberger equations without  
assuming that the anisotropy is weak. In zero
field,   all quantities are coordinate independent; besides, as
$T\to T_c$, $g \to 1$. Then, Eq. (\ref{eil1}) gives
 \begin{equation}
f ={1 \over \hbar\omega^{\prime\,}}\Big(\Delta 
+\frac{\langle\Delta\rangle}{2\omega
\tau}\Big)\equiv\frac{D}{\hbar\omega^{\prime\,}}\,,\label{imp.3}  
\end{equation}
where $\omega^{\prime\,}=\omega+1/2\tau$. 
 Substitute this   in Eq. 
(\ref{gap1}) to obtain:
\begin{equation}
\ln\frac{T_{c0}}{T_c}=\frac{\pi T_c}{\hbar\tau}\,(1- 
\langle\Omega\rangle^2)  
\sum_{\omega >0}{1\over\omega\omega^{\prime\,}}  \,.\label{Tc}
\end{equation} 
Hence $T_c=T_{c0}$ for $\tau\to\infty$ and any gap anisotropy; the same is
true for the isotropic gap ($\Omega =1$) and any $\tau$. 
This equation can be written as 
\begin{equation}
\ln\frac{T_{c0}}{T_c}=(1- \langle\Omega\rangle^2) \Big[\psi\Big({1+\mu
 \over 2}\Big)-\psi\Big({1\over 2}\Big)\Big]\,, 
 \label{digamma} 
\end{equation}
where $\mu= \hbar/ 2\pi T_c\tau $ and $\psi$ is the di-gamma function. For
a weak anisotropy $\langle\Omega\rangle^2=1-\chi$ with $\chi\ll 1$, this
reduces to Hohenberg's result. \cite{Hoh}  
Formally, Eq. (\ref{digamma}) is reminiscent of the case of magnetic
impurities; the factor $1-\langle\Omega\rangle^2$,
however, makes a difference. For $\mu \ll 1$, one has 
\begin{equation}
T_c=T_{c0}-\frac{\pi\hbar}{8\tau}(1- 
\langle\Omega\rangle^2) \,.\label{near_clean} 
\end{equation}
 For large $\mu$'s, unlike the case of the magnetic pair-breaking,  we
obtain
\begin{equation}
T_c=T_{c0}\,[\Delta_0(0) \tau/\hbar]^{
\langle\Omega\rangle^{-2}-1} \,,\label{near_dirty} 
\end{equation}
where $\Delta_0(0)=1.76\, T_{c0}$. Therefore, $T_c$ does
not turn zero at a finite $\tau$, unless $\langle\Omega\rangle=0$ as, e.g,
for the d-wave superconductors.\cite{PP} 
\\ 
 
{\it Anisotropy near $T_c$.} 
As is seen from Eq. (\ref{imp.3}), impurities cause isotropization of
amplitudes $f$, and one expects the macroscopic anisotropy 
to be suppressed by scattering. To address this question, one has to derive
the GL equations in the presence of impurities following
basically the work \cite{Gorkov} for {\it clean} superconductors. As
mentioned above, the same mass tensor enters both the first and the second 
GL equations.  We focus on the current equation  because this is
  an easier task.\cite{rem} Within Eilenberger formalism this is done in the
clean case by expanding $f$ near $T_c$ in two small parameters:
$\Delta/\hbar\omega\sim\sqrt{\delta t}$ and 
${\bf v} {\bf \Pi}\Delta /\hbar\omega^2\sim  \xi_0\Delta/\xi T_c \sim 
\delta t$ (here $\delta t=(T_c-T)/T_c$ and $\xi_0$ is zero-$T$ coherence
length):
\begin{equation}
f =\frac{\Delta }{\hbar\omega}+a \,\frac{{\bf v} {\bf
\Pi}\Delta }{\hbar\omega^2}+ {\cal O}(\delta t^{3/2})\,.
\label{series}
\end{equation}
Substituting this in Eq. (\ref{eil1}) one obtains $a =-1/2$. 
The second GL equation follows by using Eq. (\ref{eil5}) in which we
substitute $g\approx 1-ff^+/2$   with $f$'s of Eq. (\ref{series}):
\begin{equation}
j_i=-\frac{7\zeta(3)|e|\hbar N(0)}{4\pi^2T_{c0}^2 }\,\langle
\Omega^2v_iv_k\rangle
\,{\rm Im}\Psi^*\Pi_k\Psi\,.
\label{j_clean}
\end{equation}
In the London limit $\Psi = \Psi_0 e^{i\theta}$ with a constant $\Psi_0$,
and
\begin{equation}
j_i=-\frac{c\phi_0}{4\pi^2 }\,(\lambda^2)^{-1}_{ik} \Big(\nabla
\theta+\frac{2\pi}{\phi_0}{\bf A}\Big)_k\,,
\label{j_clean}
\end{equation}
with
\begin{equation}
(\lambda^2)^{-1}_{ik}=\frac{14\zeta(3)e^2  N(0)}{\pi c^2 T_{c0}
}\,\Psi_0^2\langle \Omega^2v_iv_k\rangle\,.
\label{lambda_clean}
\end{equation}
The anisotropy parameter follows:  
\begin{equation}
\gamma ^2(T_c) = \frac{\lambda^2_{cc}}{\lambda^2_{aa}}=
\frac{\langle \Omega^2 v_a^2\rangle }{\langle \Omega^2 v_c^2\rangle}\,.
\label{anis_clean}
\end{equation}
In fact, this is the result of Ref. \onlinecite{Gorkov}. 
 
In the presence of impurities, the first order term in the expansion
(\ref{series}) should have the form (\ref{imp.3}). With $D$ defined in Eq.
(\ref{imp.3}), we  verify readily that 
\begin{equation}
f =\frac{D }{\hbar\omega^{\prime\,}}-\frac{{\bf v} {\bf
\Pi}D }{2\hbar\omega^{\prime\,\,2}} +{\cal O}(\delta
t^{3/2})\, 
\label{series1}
\end{equation}
  satisfies Eq. (\ref{eil1}). Writing $D$ in the form:
\begin{equation}
D =\Psi \Big(\Omega + \frac{\langle\Omega\rangle}{2\omega
\tau}\Big)\equiv \Psi\,\Omega^{\prime\,}\,,
\label{Omega'}
\end{equation}
we obtain with the help of Eqs. (\ref{eil5}) and (\ref{series1}): 
\begin{equation}
j_i=-\frac{2\pi |e|  N(0)T}{\hbar^2} \sum_{\omega}  
\frac{\langle\Omega^{\prime\, 2} v_iv_k \rangle}{\omega^{\prime\, 3}}
\,{\rm Im}\Psi^*\Pi_k\Psi\,.
\label{j_general}
\end{equation}
In the London limit, we have:
\begin{equation}
(\lambda^2)^{-1}_{ik}=\frac{ 16\pi^2 e^2N(0)T_c }{c^2\hbar^3}\,\Psi_0^2
\sum_{\omega}\frac{\langle\Omega^{\prime\, 2} v_iv_k \rangle}
{\omega^{\prime\,3}}\,,
\label{lambda_general}
\end{equation}
and 
\begin{equation}
\gamma ^2(T_c)=\frac{\lambda^2_{cc}}{\lambda^2_{aa}}=\frac
{\sum_{\omega}\langle \Omega^{\prime\, 2} v_a^2 \rangle/\omega^{\prime\, 3}}
{\sum_{\omega}\langle \Omega^{\prime\, 2}v_c^2 \rangle/\omega^{\prime\, 3}}\,.
\label{Pokr}
\end{equation}
This is the result of Ref. \onlinecite{PP}. In the clean limit it reduces
to Eq. (\ref{anis_clean}), whereas  the effect of impurities  on $\gamma$
  depends on the order parameter symmetry.

For the d-wave symmetry $\langle\Delta\rangle =0$ and
$\Omega^{\prime\,}=\Omega $. In other words, the strong $T_c$ suppression
notwithstanding, non-magnetic impurities do not affect 
$\gamma$. For order parameters with a non-zero
$\langle\Delta\rangle$, the strong scattering  erases the effect of  gap
anisotropy on $\gamma$ altogether:
 \begin{equation}
\gamma_{\rm dirty}^2 =  
\frac{ \langle v_a^2\rangle}{ \langle v_c^2\rangle}
\, .\label{gam_lam_dirty}
\end{equation}
Hence, in the dirty limit, all parts of the Fermi surface contribute evenly 
 to the anisotropy parameter as is the case for isotropic gaps.    \\

{\it T dependence of $\gamma=\lambda_c/\lambda_{a}$.} 
To address this question in the full temperature range 
one has to study weak supercurrents, i.e.,  turn to Eq. (\ref{eil5}). 
We consider only the clean case for which $f_0             ,g_0             $ in
the absence of currents are:
\begin{equation}
f _0 =f ^+_0={\Delta_0 \over
\beta },\quad g _0  =  {\hbar\omega\over
\beta } ,\quad
\beta ^2=\Delta_0 ^2+ \hbar^2\omega^2\,; \label{f_0}\\
\end{equation}
 in general, both $\Delta_0$ and $\beta$ depend  on
${\bf k}_F$ . A weak supercurrent causes
the order parameter
$\Delta$ and  the amplitudes $f$ to acquire an overall phase 
$\theta({\bf r})$. We look for the  perturbed solutions in the form:  
\begin{eqnarray}
\Delta  = \Delta _0 \, e^{i\theta},\,\,\,\,\,
f =(f_0  +f_1)\,e^{i\theta},\nonumber\\ 
f^{+} =(f_0  +f_1^+ )e^{-i\theta},\,\,\,\,\,
g =g_0 +g_1             , 
\label{eq67}
\end{eqnarray}
where the subscript 1 denotes  corrections.  In the London 
limit, the only coordinate dependence is that of the phase $\theta$, i.e.,
  $f_1 ,g_1 $ can be taken as ${\bf r}$
independent. \cite{rem1}  Equations (\ref{eil1}-\ref{eil3}) then give:
 \begin{eqnarray}
\Delta_0 g_1 &-&\hbar\omega f_1 =i\hbar f_0  {\bf v}{\bf P}/2\,,\nonumber\\ 
\Delta_0 g_1 &-&\hbar\omega f^+_1=i\hbar f_0 {\bf
v}{\bf P}/2\,,\label{e38}\\ 
2g_0 g_1 &=&-f_0 (f_1 +f^+_1 ) \,.\nonumber
\end{eqnarray}
Here,  
${\bf P}= \nabla\theta+ 2\pi{\bf A}/\phi_0\equiv 2\pi\, {\bf a}/\phi_0$.
To evaluate the current (\ref{eil5}), one solves the system (\ref{e38}) for
$g _1 $:
 \begin{equation}
g _1 =i\hbar\,\frac{\Delta_0 ^ 2 }{2\beta ^3}\, 
 {\bf v}{\bf P}  \,.
\label{g_corr}
\end{equation}
Then one obtains the London relation between the current and the vector
potential,
$4\pi j_i/c=-  (\lambda^2)_{ik}^{-1} a_k$, with 
\begin{equation}
(\lambda^2)_{ik}^{-1}= \frac{16\pi^2 e^2T}{ 
c^2}\,N(0)  \sum_{\omega} \Big\langle\frac{
\Delta_0^2v_iv_k}{\beta ^{3}}\Big\rangle \,.  \label{e31}
\end{equation}

The anisotropy parameter now reads:
\begin{equation}
\gamma^2=\frac{\lambda^2_{cc}}{\lambda^2_{aa}}= \frac{  
 \langle  v_a^2\,\Delta_0^2\sum_{\omega}\beta ^{-3}\rangle  
}{ \langle  
v_c^2\,\Delta_0^2\sum_{\omega}\beta ^{-3}\rangle }
 \,. \label{gamma(T)}
\end{equation}

As $T\to 0$, we have $2\pi
T\Delta_0^2 \sum_{\omega}\beta ^{-3}\to 1$, and
\begin{equation}
\gamma^2(0)=  \frac{  \langle v_a^2\rangle}{\langle v_c^2\rangle}\,.  
\label{gamma(0)}
\end{equation}
Note that the gap and its anisotropy 
do not enter this result. The physical reason for
this is in the Galilean invariance of the superfluid flow in the absence of
scattering: all charged particles take part in the supercurrent.\cite{VP}

 Near $T_c$, $\sum_{\omega}\beta ^{-3} \to  
    7\zeta(3)/8\pi^3T_c^3$, and  we obtain the GL result
(\ref{anis_clean}) which amplifies contribution of 
 the Fermi surface pieces with large gap to the parameter
$\gamma$. Thus,  the anisotropy parameter depends on 
$T$, the feature absent in  superconductors with isotropic gaps.

It is worth noting that in literature the superconducting anisotropy
is commonly referred to as the ratio $H_{c2,a}/H_{c2,c}$, an important
figure for applications, but a difficult quantity to evaluate  
for anisotropic Fermi surfaces, not to speak about anisotropic gaps. On the
other hand, measurements of the $\lambda$-anisotropy such as, e.g., the
high-field torque technique is more demanding than the resistive
determination of $H_{c2 }$ since one has to work in the reversible domain.
One should also note that, theoretically, the ratios of $H_{c2 }$'s and of
$\lambda$'s are not necessarily the same, except near $T_c$ where their
equality   is provided by the GL theory. \\

   It is of interest to examine the consequences of our results   for  {\it
MgB$_2$.} The reported $\gamma$'s vary from 1.7 to 8,
\cite{Lima,madison,Simon,BKC} or even higher as in Ref.
\onlinecite{Shinde}.  In all these reports, different techniques for
extracting the anisotropy and  samples with different resistivity ratios
were used. 

 Consider a model material with the gap anisotropy given by
\begin{equation}
\Omega= \Omega_{1,2}\,,\quad {\bf v}\in   F_{1,2} \,, 
 \label{e50} 
\end{equation}
where $F_1,F_2$ are two sheets of the Fermi surface.  Denoting
the densities of states on the two parts as $N_{1,2}$, and assuming the
quantity $X$ being constant at each sheet, we obtain for the general
averaging (\ref{norm}): 
\begin{equation}
\langle X \rangle = (X_1 N_1+X_2 N_2)/N(0) =  \nu_1X_1+
\nu_2X_2\,,\label{norm2}
\end{equation}
where we introduce normalized densities of state
$\nu_{1,2}= N_{1,2}/N(0)$ for brevity.  We have then instead of Eq.
(\ref{norm}): 
\begin{equation}
\Omega_1^2 \nu_1+\Omega_2^2\nu_2=1\,,\quad \nu_1+\nu_2=1\,.\label{norm1}
\end{equation}
  We also assume that
the two parts of the Fermi surface have the symmetries of the total, e.g.,
$\langle {\bf v}\rangle_{1}=0$ where the average is performed only over the
first Fermi sheet. Within this model, Eq. (\ref{gamma(T)}) reads: 
\begin{equation}
\gamma^2=  \frac{\sum_i\nu_i\Omega_i^2   
 \langle  v_a^2\rangle_i\, \sum_{\omega}\beta_i^{-3}
 }{\sum_i\nu_i\Omega_i^2   
 \langle  v_c^2\rangle_i\, \sum_{\omega}\beta_i^{-3}}
  \,,\quad i=1,2\,, \label{gamma(T_2)}
\end{equation}
where $\beta_i=\sqrt{\hbar^2\omega^2+\psi^2(T)\Omega^2_i}\,$. 

Based on the band structure calculations, the relative densities of states 
$\nu_1$ and $\nu_2$ of our model are $\approx\,$0.56 and 0.44.
\cite{Bel,Choi} The ratio $\Delta_2/\Delta_1=\Omega_2/\Omega_1\approx
4$, if one takes the averages of $6.8\,$ and $1.7\,$meV 
for the two groups of distributed gaps as calculated in Ref.
\onlinecite{Choi}. Then, the normalization (\ref{norm1})  yields
$\Omega_1=0.36$  and  $\Omega_2=1.45$. 

\begin{figure}
\epsfxsize .8\hsize
\centerline{
\epsffile{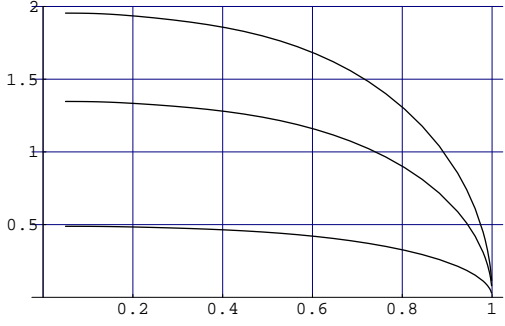}} 
\vskip .5\baselineskip
\caption{Temperature dependence of the two gaps
calculated  as $\Delta_i=\psi(T)\Omega_i$ versus  
$T/T_c$. The upper curve is
$\Delta_2/T_c$, the lower one is $\Delta_1/T_c$, and the middle curve is
$\psi(T)/T_c$ evaluated as described in the text.}
\label{f1}
\end{figure}

Now, we have all  parameters needed to solve the self-consistency equation
 (\ref{gap1}) for $\psi(T)$ with $f=\Delta/\beta$ (the clean case). This
is done numerically and the result   is shown in Fig.
\ref{f1} along with $\Delta_i(T)$.  

\begin{figure}
\epsfxsize .8\hsize
\centerline{
\epsffile{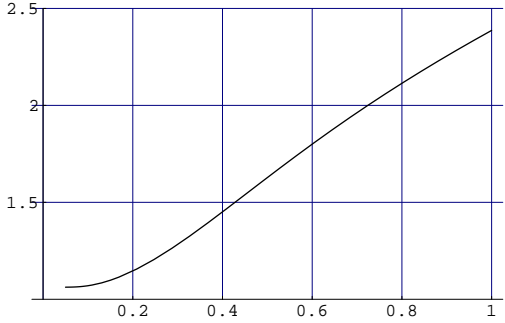}} 
\vskip .5\baselineskip
\caption{Temperature dependence of $\gamma=\lambda_c/\lambda_a$ versus 
  $T/T_c$ for clean
MgB$_2$ calculated with parameters given in the text.}
\label{f2}
\end{figure}

To evaluate $\gamma(T)$ of Eq. (\ref{gamma(T_2)}) we use the averages over
separate Fermi sheets calculated in Ref. \onlinecite{Bel}: 
$\langle v_a^2\rangle_1=33.2$, $\langle v_c^2\rangle_1=42.2$, 
$\langle v_a^2\rangle_2=23$,
and $\langle v_c^2\rangle_2=0.5\times 10^{14}\,$cm$^2$/s$^2$
The result of numerical evaluation of $\gamma(T)$ is shown in Fig.
\ref{f2}.

The ratio of $\lambda$'s can be obtained, e.g., from the
angular dependence of the reversible torque   on single crystals in
intermediate magnetic fields tilted relative to the principal crystal
directions.\cite{K88,Farrell} Some torque data for MgB$_2$ were reported by
M. Andst {\it et al.},\cite{Angst} but the $T$ dependence was not examined
in detail. The ratio $H_{c2,ab}/H_{c2,c} $ was shown to drop with
increasing $T$ from about 6 at 15 K to 2.8 at 35 K (see also Refs.
\onlinecite{Sergey,Argonne}). The authors estimate this ratio as $ \approx
2.3-2.7$ at $T_c$. Near $T_c$, the ratios of $H_{c2}$'s  should coincide
with the ratio of $\lambda$'s. In this work we estimate the latter as 
$\approx 2.4$, see Fig. \ref{f2}. Given this agreement and   the prediction
made here that the
$T$ dependence of $\gamma=\lambda_{c}/\lambda_{ab}$ should be opposite to
that of the observed behavior of $H_{c2,ab}/H_{c2,c}$, the detailed
  studies of $\gamma$  by torque or other methods are desirable.\\  

The author is grateful to L.  Gor'kov and V. Pokrovsky for informative
and helpful discussions, to K. Belashchenko and V. Antropov for providing 
 data on Fermi surface averages,  to S. Bud'ko and P.
Canfield for the interest in this work. 
This work was supported by the Director of Energy Research, Office of Basic
Energy  Sciences, U. S. Department of Energy.

 \end{multicols} 
\end{document}